# Near the Runaway: The Climate and Habitability of Teegarden's Star b

Ryan Boukrouche[1,2] , Rodrigo Caballero[1] , and Neil T. Lewis[3]
[1] Department of Meteorology, Stockholm University, Sweden; ryan.boukrouche@astro.su.se
[2] Department of Astronomy, Stockholm University, Sweden
[3] Department of Physics and Astronomy, University of Exeter, UK

## Abstract

Teegarden's Star b, a nearby terrestrial world receiving an Earth-like instellation, is a prime candidate for next-generation observatories targeting temperate exoplanets in their habitable zones. We employ a suite of three-dimensional global climate model simulations to (1) map the inner boundary of the habitable zone of Teegarden's Star b and (2) characterize its surface climate under the assumption of an Earth-analog atmosphere. Our simulations show that, with its most recently estimated instellation of 1481 W m$^{-2}$, Teegarden's Star b remains below the runaway greenhouse threshold for both low ($\alpha_s = 0.07$, ocean-dominated) and moderate ($\alpha_s = 0.30$, land-dominated) surface albedos. However, a different estimate of 1565 W m$^{-2}$ places it beyond the runaway threshold. The result that Teegarden's Star b is habitable under the most recent instellation measurement reinforces its status as one of the most compelling targets for future habitability and biosignature searches. Given the planet's proximity to the runaway threshold, it would benefit from a comparative study done with other models using different parameterizations.

*Unified Astronomy Thesaurus concepts:* Extrasolar rocky planets (511); Exoplanet atmospheres (487); Astrobiology (74); Habitable planets (695)

## 1. Introduction

Teegarden's Star is an M dwarf 12.5 lt-yr away discovered by B. J. Teegarden et al. (2003). Using the radial velocity method, three Earth-mass planets were detected around it (M. Zechmeister et al. 2019; S. Dreizler et al. 2024). This introduced some of the closest Earth analogs to date, and as such, they have been the subject of growing interest (K. G. Stassun et al. 2019; M. Micho et al. 2025). They do not transit, which is why detailed characterization has not yet been performed. It will require the capabilities of the next generation of observatories including LIFE (S. P. Quanz et al. 2022) and Extremely Large Telescope - Planetary Camera and Spectrograph (ELT-PCS; M. Kasper et al. 2021).

Teegarden's Star is especially promising for habitability searches, as it is considered relatively quiescent compared to other stars of its type (S. Dreizler et al. 2024). Therefore, flaring and harmful far-ultraviolet and extreme ultraviolet radiation affecting the atmosphere and surface conditions is less of a concern. M. Zechmeister et al. (2019) and S. Dreizler et al. (2024) suggest that planet b around Teegarden's Star may be in the optimistic habitable zone as defined by R. K. Kopparapu et al. (2014), based on its orbital and stellar parameters. Also, A. Wandel & L. Tal-Or (2019) found, using a simplified analytical model derived by A. Wandel (2018), that there are ranges of heat redistribution efficiencies that could allow for the presence of surface liquid water in high-latitude regions of Teegarden b, with surface temperature ranging from 310 to 373 K, although the corresponding substellar surface temperatures range from 340 to 400 K.

In this study, we build on previous work by using a 3D general circulation model to study the climate of Teegarden's Star b. We assess the habitability of planet b and explore the location of the inner edge of the planet's habitable zone, under the assumption that it is tidally locked and has an Earth-like atmospheric composition. In Section 2, we describe the global climate model (GCM) and the parameters used for the modeling. Our results are shown in Section 3. Finally, we summarize our conclusions and discuss the significance of our results in Section 4.

## 2. Methods

To simulate the climate of Teegarden's star b we use Isca (G. K. Vallis et al. 2018), a flexible framework for modeling the global circulation of planetary atmospheres, developed and maintained at the University of Exeter. It has been used previously to simulate the climate and atmospheric dynamics of tidally locked exoplanets (J. Penn & G. K. Vallis 2018; N. T. Lewis & M. Hammond 2022). In the present study, Isca is configured as an idealized aquaplanet general circulation model using a simplified representation of moist physics that follows D. M. Frierson (2007) and P. A. O'Gorman & T. Schneider (2008) but featuring a simple diagnostic cloud scheme (Simcloud; Q. Liu et al. 2020) and a correlated-$k$ radiative transfer code (SOCRATES; J. M. Edwards & A. Slingo 1996). A full description of the model configuration we use is included in Appendix A.

Our simulations are performed assuming that Teegarden's star b is tidally locked. We use the orbital and bulk parameters estimated for planet b around Teegarden's Star, namely a radius of 1.02 $R_{\mathrm{Earth}}$, a mass of 1.16 $M_{\mathrm{Earth}}$, a semimajor axis of 0.0259 au, and an orbital period of 4.90634 days (S. Dreizler et al. 2024). For our fiducial simulation, we further assume an instellation of 1481 W m$^{-2}$ based on the most recent estimates reported by S. Dreizler et al. (2024). We use a BT-Settl spectrum corresponding to Teegarden's star in the radiative transfer calculation. To map the inner edge of the habitable zone, we run a series of additional experiments where the instellation is varied. We consider two values for the surface albedo, 0.07 and 0.3, which are intended to approximate ocean- and land-dominated surfaces, respectively.

For all simulations we assume a present-day Earth-like composition of 78.084% $N_2$ and 20.947% $O_2$. The $CO_2$ concentration is 400 ppmv. Trace gases are included with the following abundances: 1 ppmv $CH_4$, 0.001 ppmv CO, and 0.03 ppmv $H_2$. Our simulations do not include ozone; however, we ran extra simulations that do include an Earth-like ozone distribution and found our main results to be unchanged. Water vapor is included in the model and advected by the atmospheric circulation. Its overall abundance is determined by a balance between precipitation and evaporation.

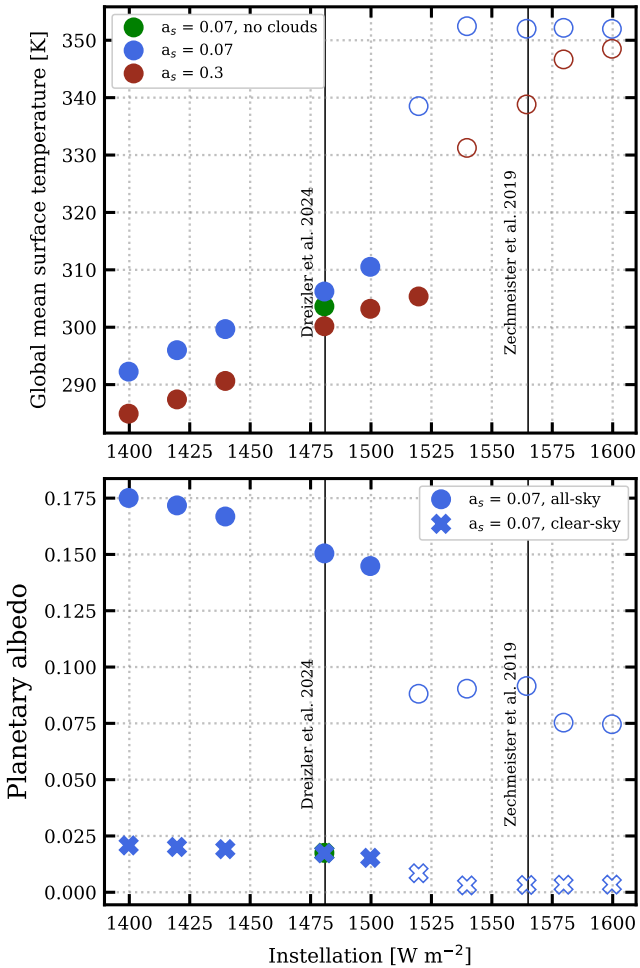

**Figure 1.** Global mean surface temperature and planetary albedo as a function of instellation for all simulations. The markers without filling represent the runaway cases shown over the last 2 yr of the simulation. The clear-sky albedo values are computed using the clear-sky fluxes of the same simulations. The green markers correspond to a separate simulation with $a_s = 0.07$, where clouds were turned off.

All simulations are run for several Earth years, from 8 to 41, with diagnostics reported over the final steady-state period or the final few months in our runaway cases. For each stable climate, we have checked to ensure that the top-of-atmosphere energy budget and the globally averaged water budget are closed before the start of the diagnostic period.

### 3. Results

Teegarden b orbits at about 0.0259 au from its host star according to the most recent measurement (S. Dreizler et al. 2024), yielding a stellar constant of 1481 W m$^{-2}$. We explore a range of instellation values between 1400 and 1600 W m$^{-2}$ to estimate where the planet features a stable climate. Figure 1(a) shows the globally averaged $T_s$ obtained in each simulation as a function of instellation, $S$. An increase in instellation leads to a higher equilibrium surface temperature, as the system adjusts to restore radiative balance. Simulations that reach a stable equilibrium climate are indicated with filled markers, while those that enter the runaway greenhouse transition are indicated with empty markers. For a surface albedo of 0.07, we find that stable climate states are obtained for $S \leqslant 1500$ W m$^{-2}$, while for a surface albedo of 0.3, stable climates are obtained for $S \leqslant 1520$ W m$^{-2}$. The trend beyond this point becomes flatter because beyond the onset of the runaway, the air temperature and specific humidity exceed the nominal range that the model can account for, and it crashes. In reality, the trend would continue to increase up to a point that is yet to be explored with models that can simulate high-temperature and potentially nondilute atmospheres. Time series of the difference between the absorbed stellar radiation (ASR) and the outgoing longwave radiation (OLR) are shown in Figure 2, for four cases either side of the runaway transition. For those beyond the transition, enhanced atmospheric water vapor content inhibits cooling to space, which causes the ASR and OLR to diverge, so that a stable climate is not obtained. Our model is not adapted to simulate the postrunaway state (e.g., R. Boukrouche et al. 2021; G. Chaverot et al. 2023), so instead, the simulations are run until they crash when the model becomes too hot.

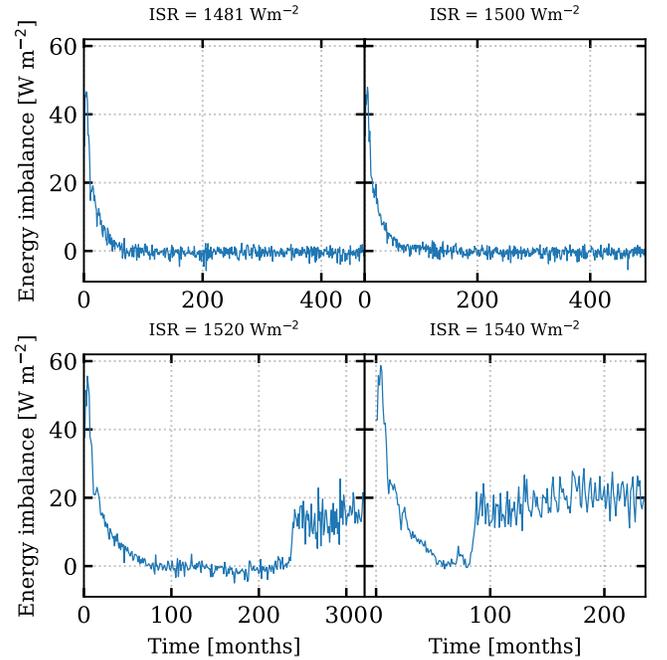

**Figure 2.** Time series of the difference between ASR and OLR across the runaway greenhouse instellation threshold with a surface albedo of 0.07.

Figure 1(b) shows the global mean clear-sky and all-sky (i.e., cloudy) planetary albedo as a function of instellation of all simulations for a surface albedo of 0.07. The case with $a_s = 0.3$ yields the same trend. For stable climates, the planetary albedo is greater than the surface albedo because of the presence of clouds on the planet's dayside, whose effect on the albedo can be seen from the difference with the clear-sky albedo. The reflection of incoming radiation by clouds decreases as the instellation is increased, which accelerates the runaway transition. Inspection of the distribution of cloud water content (not shown) reveals that the downward trend in the all-sky planetary albedo is due to the dayside cloud structure being lifted upward and becoming optically thinner. The clear-sky albedo and the all-sky albedo for the runaway cases are lower than or close to the surface albedo value. This is due to atmospheric absorption of incoming solar radiation, which is enhanced for planets orbiting M dwarfs (compared to, e.g., the Sun) because the stellar spectrum is shifted toward longer wavelengths. This effect hampers the habitability of planets orbiting M dwarfs (J. Yang et al. 2013).

**Table 1**
Surface Temperature $T_{\mathrm{surf}}$ and Planetary Albedo $a_p$

| Surface Albedo | 0.07 | 0.3 |
| --- | --- | --- |
| Global mean $T_{\mathrm{surf}}$ [K] | 306 | 300 |
| Maximum $T_{\mathrm{surf}}$ [K] | 339 | 334 |
| Minimum $T_{\mathrm{surf}}$ [K] | 285 | 271 |
| Dayside mean $a_p$ | 0.267 | 0.286 |

Taking the most recent estimate of the installation received at Teegarden's Star b (1481 W m$^{-2}$; S. Dreizler et al. 2024) to be correct, our simulations suggest that for an Earth-like surface pressure and atmospheric composition, Teegarden b is in its habitable zone but lies very close to the inner edge. This result is robust to variations in the choice of surface albedo. Moreover, the results presented here are unchanged if we include an Earth-like ozone layer in the simulation. However, if instead of 1481 W m$^{-2}$, the installation is closer to 1565 W m$^{-2}$ (as estimated by M. Zechmeister et al. 2019), then the planet would be beyond the runaway greenhouse threshold. Figure 3 (top row) shows the surface temperature distribution for the two simulations run with an installation of 1481 W m$^{-2}$, and Table 1 lists the global mean, minimum, and maximum values for $T_{\mathrm{surf}}$, as well as the planetary albedo, for these cases. Using this installation and an albedo of 0.07, the global mean surface temperature is 306 K, about 18 K higher than on present-day Earth, while the maximum temperature is 339 K. While the boiling point of water is never exceeded, so that the planet remains "habitable," we note that temperatures this high would be very uncomfortable or fatal for life adapted for Earth (see, e.g., the 50°C contour in Figure 3). The day–night temperature difference within 50° of latitude is about 49 K with $a_p = 0.07$ and 52 K with $a_p = 0.3$, suggesting that the planet might be relatively far from a weak temperature gradient regime, consistent with the planet's relatively rapid rotation rate ($P_{\mathrm{orb}} = 4.9$ days).

The second and third rows of Figure 3 show the distribution of surface evaporation and precipitation. For both cases, evaporation occurs mainly on the dayside, but precipitation falls on the nightside. This occurs because, in our simulations, convection is inhibited on the dayside due to enhanced absorption of incoming stellar radiation (associated with the M dwarf stellar spectrum). In particular, annual precipitation in most regions of the dayside subtropics is on the order of 163 mm for $a_s = 0.07$ and 261 mm for $a_s = 0.3$, which is comparable to the range observed in the wettest regions of the Sahara on Earth (M. Armon et al. 2024). We note that our model is configured so that the surface acts as an infinite reservoir of water, which means that evaporation can always occur. Our simulations suggest that the "real" Teegarden b would likely have a dry dayside (unless it is an ocean world, or there is an efficient return flow of water from the nightside to the dayside). This would reduce evaporative cooling, causing dayside surface temperatures to be enhanced relative to those obtained in our simulations (N. T. Lewis et al. 2018). We show the same quantities in a near-runaway case at incoming stellar

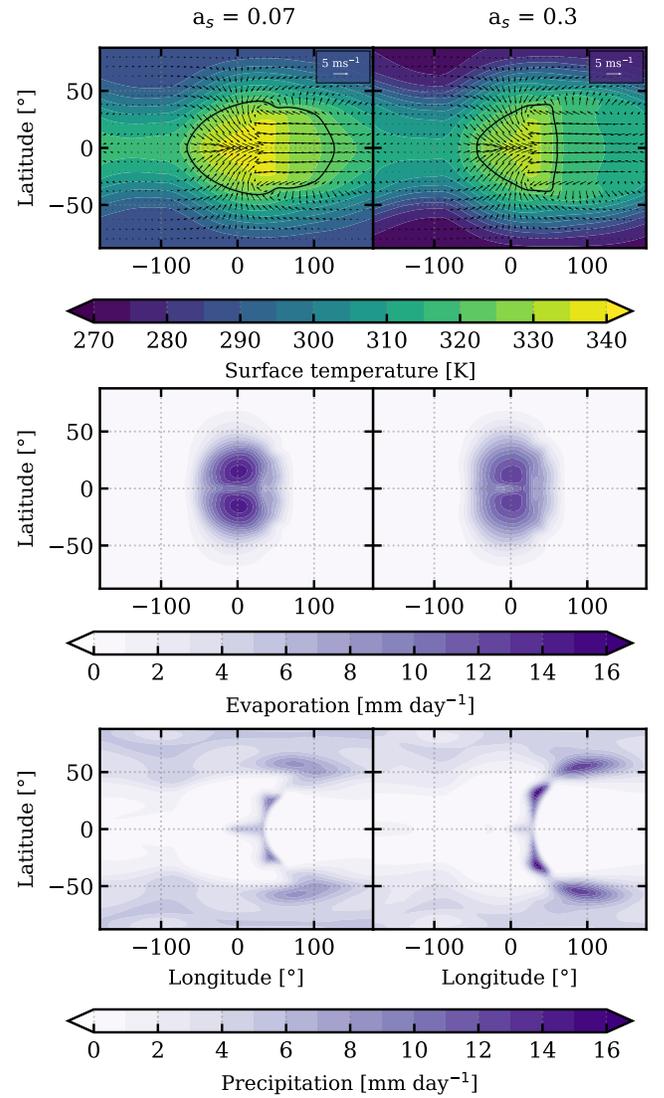

**Figure 3.** Latitude–longitude maps of surface temperature overlaid with the wind field at the bottom layer, and surface precipitation. The black contours indicate where the surface temperature is 50° Celsius.

radiation (ISR) = 1580 W m$^{-2}$ in the Appendix, in Figure 5 in Appendix C.

## 4. Discussion

We show that for an installation of 1481 W m$^{-2}$ (S. Dreizler et al. 2024), the GCM simulations presented here suggest Teegarden's Star b is habitable. This conclusion holds across a range of surface albedos from water-dominated to land-dominated and whether or not an Earth-like ozone layer is accounted for. However, it is only 20–40 W m$^{-2}$ from the runaway threshold. The inner edge found in this work stands at $1510 \pm 10$ W m$^{-2}$ with a surface albedo of 0.07 and $1530 \pm 10$ W m$^{-2}$ with a surface albedo of 0.3, in other words, at a distance of [0.02574, 0.02557] au and [0.02557, 0.0254] au, respectively. Though the sensitivity to surface albedo is therefore relatively small, it illustrates that the habitable zone does depend on factors intrinsic to the planet, not just on orbital parameters.

We currently have two measurements for the orbital distance of Teegarden b from which installations can be derived. The other is 1565 W m$^{-2}$ from M. Zechmeister et al. (2019). So far, there has been no clear refutation of this latter result, which motivates us to take it as a serious possibility. Using this instellation value, we find that the planet would undergo a runaway greenhouse effect, which would desiccate the planet and prevent the scenario described by A. Wandel & L. Tal-Or (2019), which found that some regions of the planet might be cold enough for liquid water to exist.

We note that our conclusions will be sensitive to the choices we have made in our study design. First, the assumption of a 1 bar $N_2$ atmosphere is relatively strong, since the origins of nitrogen on Earth are not yet fully understood, and hence, we cannot yet predict its presence and abundance on exoplanets (Y. Li 2024). However, its detection, although very challenging, might be possible using indirect means, notably with collisional pairs $N_2$–$N_2$, which have an absorption feature at 4.15 $\mu$m (E. W. Schwieterman et al. 2015). This is close to the short-wave boundary of the spectral range currently envisioned for LIFE, 4 $\mu$m. The $CO_2$ abundance of 400 ppm is also a strong assumption, although it is partially regulated by the carbonate-silicate feedback (J. C. G. Walker et al. 1981; R. Graham & R. Pierrehumbert 2024), which carries its own set of assumptions about the planet's tectonic regime and volcanic activity. LIFE may be able to constrain the planet's $CO_2$ abundance using its absorption feature around 15 $\mu$m.

More generally, our conclusions will also be sensitive to the specific model configuration we have used. In particular, our model uses a relatively simple convective adjustment scheme; it is plausible that the extent of convective suppression due to the M-star stellar spectrum is sensitive to this choice. Simulations of Trappist-1e presented by D. E. Sergeev et al. (2022b) show that different choices for the convective adjustment scheme in the Met Office Unified Model GCM can lead to different climate states. Comparison projects such as the Trappist-1 Habitable Atmosphere Intercomparison (THAI; T. J. Fauchez et al. 2021; M. Turbet et al. 2022; D. E. Sergeev et al. 2022a) have shown that different GCMs configured to simulate the same planet can produce a range of climates and circulation regimes (presumably owing to differences between the parameterizations included in each model). For example, models capable of consistently simulating nondilute atmospheres may explore the possibility that under a range of instellation values, the planet's atmosphere might be in a moist greenhouse state (J. F. Kasting et al. 1984) instead of a runaway, where water builds up enough that the stratosphere becomes moist, driving photodissociation and loss of water to space. Given Teegarden's Star b's prospects for habitability and future characterization, and our finding that it resides close to the inner edge of the habitable zone, we suggest that it would benefit from a detailed model inter-comparison similar to THAI.

Teegarden's Star b has long been presumed to be one of our most promising Earth analogs. Our simulations suggest that this presumption is still warranted, although its habitability is highly dependent on the orbital distance measured. The difference between the two orbital radius estimates currently available is large enough to prevent a robust estimation of the actual instellation of the planet, and future measurements are still needed to help constrain it further. Teegarden b remains a promising target to follow up with future direct imaging observatories aimed at potentially habitable worlds like LIFE, whether it turns out to be a habitable planet or a postrunaway planet.

## Acknowledgments

The authors thank Prof. Markus Janson and Prof. Thorsten Mauristen for useful insights, and Dr. Daniel Williams for technical support. We thank the reviewer for feedback, which helped improve this Letter.

These computations were enabled by resources provided by the National Academic Infrastructure for Supercomputing in Sweden (NAISS), partially funded by the Swedish Research Council through grant agreement No. 2022-06725. This work was also supported by an interdisciplinary postdoctoral fellowship issued by the Section for Mathematics and Physics at Stockholm University. N.T.L. is supported by STFC grant ST/Y002156/1.

*Software:* NUMPY (C. R. Harris et al. 2020), SCIPY (E. Jones et al. 2001), MATPLOTLIB (J. D. Hunter 2007), SOCRATES (J. M. Edwards & A. Slingo 1996), XARRAY (S. Hoyer & J. Hamman 2017).

## Appendix A
## Model Description

This Appendix provides a complete description of the Isca model configuration used for this study.

### A.1. Dynamical Core

The dynamical core is a spectral core developed by the Geophysical Fluid Dynamics Laboratory. The dynamical core solves the primitive equations using a pseudospectral method in the horizontal and finite differences in the vertical. Linear terms in the fluid dynamics equation are evaluated after transformation to spherical harmonics, while parameterizations and nonlinear terms are evaluated on the real-space grid. We use a T42 triangular truncation for the spherical harmonics, which yields a resolution of 2°.8 (319–12 km from equator to pole) in longitude and 2°.8–3°.4 (317–388 km from equator to pole) in latitude. We use a hybrid terrain following coordinate with 48 pressure levels defined $p_k = a_k + p_{\text{surf}} b_k$. The coefficients $a_k$ and $b_k$ are shown in Appendix B.

### A.2. Radiation

Radiation is parameterized using SOCRATES v.24.03 (J. M. Edwards & A. Slingo 1996), which solves the plane-parallel two-stream equations and accounts for Rayleigh scattering and Mie scattering of water droplets and ice crystals, which are both assumed to be spherical.

We use different opacity sources shown in Table 2. The sources in the first four rows (HITRAN 2016, HITEMP 2019 1.0, ExoMol WCCRMT 1.0, and HITRAN 2020 1.0) are mostly experimental, with theoretical calculations and semi-empirical formulae used to interpolate or extrapolate missing data. The next four (MT_CKD 3.0, ExoMol POKAZATEL 2.0, ExoMol UCL-4000 1.0, and ExoMol YT34to10 1.0) are mostly theoretical, constrained by in situ and laboratory measurements when possible. RACPPK 1.0 involves theoretical calculations only. We include the continua $H_2O$–$H_2O$, $H_2$–$CH_4$, $H_2$–$H_2$, $N_2$–$H_2$, $N_2$–$N_2$, $N_2$–$H_2O$, $O_2$–$CO_2$, $O_2$–$N_2$, $O_2$–$O_2$, $CO_2$–$CO_2$, $CO_2$–$H_2$, and $CO_2$–$CH_4$.

We use a long-wave and short-wave spectral file with 16 spectral bands from 1 to 35,000 cm$^{-1}$, with bandwidths decreasing toward shorter waves, as seen in Figure 4. Opacities are tabulated in pressure–temperature space. The pressure table

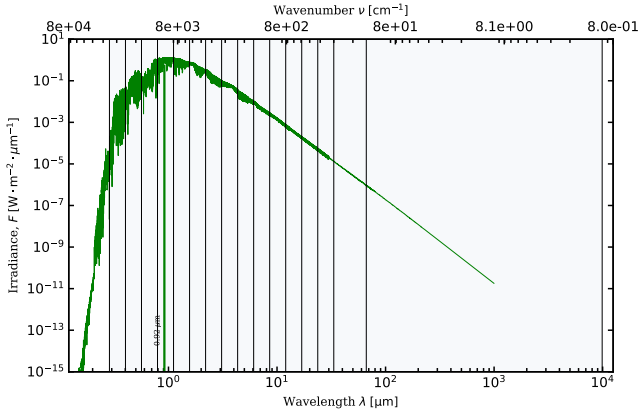

**Figure 4.** Irradiance of an spectral energy distribution (SED) with an effective temperature of 3000 K and a log(g) of 5 $\log_{10}(\mathrm{cm\,s^{-2}})$ representing Teegarden's Star. The edges of the 16 spectral bands used by SOCRATES are shown as vertical lines.

**Table 2**
Spectroscopic Data Sources

| Source | Data |
| --- | --- |
| HITRAN 2016 (I. E. Gordon et al. 2017) | All continua, except $H_2O$ |
| HITEMP 2019 1.0 (G. Li et al. 2015) | CO |
| ExoMol WCCRMT 1.0 (C. M. Western et al. 2018) | $N_2$ |
| HITRAN 2020 1.0 (I. E. Gordon et al. 2022) | $O_3$, $O_2$ |
| MT_CKD 3.0 (Github[a]) | $H_2O$ continua |
| ExoMol POKAZATEL 2.0 (O. L. Polyansky et al. 2018) | $H_2O$ |
| ExoMol UCL-4000 1.0 (X. Huang et al. 2023) | $CO_2$ |
| ExoMol YT34to10 1.0 (S. N. Yurchenko et al. 2017) | $CH_4$ |
| RACPPK 1.0 (E. Roueff et al. 2019) | $H_2$ |

**Note.**
[a] https://github.com/DavidSAmundsen/socrates_tools/tree/master/continuum/h2o

is evenly spaced in log pressure, ranging from $10^{-6}$ to 1000 bar using 80 bins. The temperature table goes from 50 to 2900 K linearly using 42 bins. The long-wave and short-wave files are distinguished by transmittance weights, which are Planckian at the transmission temperature for the former and stellar at the top of the atmosphere for the latter.

We use a stellar spectrum corresponding to Teegarden's Star taken from the BT-Settl model grid of theoretical photospheric spectra hosted by the Spanish Virtual Observatory. It uses an effective temperature of 3000 K and a log(g) of 5 $\log_{10}$ (cm $\mathrm{s}^{-2}$). For the purposes of this work, it is an acceptable proxy for Teegarden's Star, which has a relatively similar luminosity, mass, and radius (K. G. Stassun et al. 2019; E. Agol et al. 2021).

### A.3. Moist Processes

Moist processes are parameterized following D. M. Frierson (2007) and P. A. O'Gorman & T. Schneider (2008). We use the "simple Betts–Miller" convective relaxation scheme to parameterize convection. Humidity is relaxed toward a fixed relative humidity of 0.7, and the convective relaxation timescale is taken to be 2 hr. Large-scale condensation is computed according to P. A. O'Gorman & T. Schneider (2008), which is based on a stochastic model of moisture kinematics introduced by R. T. Pierrehumbert et al. (2007). Condensation occurs when the relative humidity exceeds unity, and reevaporation can occur in subsaturated layers. All precipitation is liquid; we neglect snow formation. We use a simplification where we compute the relative humidity as $R_\mathrm{H} = \frac{q_\mathrm{v} R_\mathrm{v} \max(p, e_\mathrm{sat})}{R_\mathrm{d} e_\mathrm{sat}}$ instead of $R_\mathrm{H} = \frac{q_\mathrm{v} R_\mathrm{v} \max(p - (1 - \frac{R_\mathrm{d}}{R_\mathrm{v}})e_\mathrm{sat}, e_\mathrm{sat})}{R_\mathrm{d} e_\mathrm{sat}}$, where $q_\mathrm{v}$ is the specific humidity, $e_\mathrm{sat}$ is the saturation vapor pressure, and $R_\mathrm{d}$ and $R_\mathrm{v}$ are the gas constants of dry air and vapor.

To incorporate the radiative effects of clouds into the model, use the idealized SIMCLOUD scheme described by Q. Liu et al. (2020). It is a diagnostic cloud scheme, with cloud formation determined by the local conditions in each column at each time step. There is no explicit transport of clouds by wind. Large-scale clouds are diagnosed from the relative humidity, and marine low stratus clouds are determined largely as a function of inversion strength. The cloud fraction, the effective radius of cloud droplets, and in-cloud water mixing ratio are parameterized as described in Q. Liu et al. (2020).

### A.4. Surface and Top-of-atmosphere Parameterizations

Momentum, heat, and moisture surface fluxes are computed with drag equations defined in D. M. Frierson et al. (2006), with drag coefficients computed with a simplified Monin–Obukhov similarity theory. In the mixed layer, we update the surface temperature based on the surface fluxes and the mixed layer depth of a slab ocean. The slab depth is 2.5 m, yielding a small thermal inertia that quickly adjusts the surface temperature to the bottom air temperature. The momentum, heat, and moisture roughness lengths used in the computation of the surface fluxes are respectively set to $5 \cdot 10^{-3}$, $10^{-5}$, and $10^{-5}$ m following P. A. O'Gorman & T. Schneider (2008). Vertical diffusion within the boundary layer is parameterized using a nonlocal K scheme similar to I. Troen & L. Mahrt (1986). We use a sponge at the top of the atmosphere between 1 and 20 Pa to improve the stability of the model using Rayleigh friction with a damping time of 12 hr.

### Appendix B
### Model Levels

The coefficients $b_k$ and $p_k$ defining the $\sigma$ pressure coordinates $\sigma = b_k p_\mathrm{surf} + p_k$ are shown below:

$$\begin{bmatrix}
0.00000 & 0.00000 & 0.00000 & 0.00000 & 0.00000 & 0.00000 \\
0.00000 & 0.00000 & 0.00000 & 0.00000 & 0.00000 & 0.00000 \\
0.00000 & 0.00000 & 0.00000 & 0.00000 & 0.00000 & 0.00000 \\
0.00000 & 0.00000 & 0.00000 & 0.00000 & 0.00000 & 0.00000 \\
0.00000 & 0.01253 & 0.04887 & 0.10724 & 0.18455 & 0.27461 \\
0.36914 & 0.46103 & 0.54623 & 0.62305 & 0.69099 & 0.75016 \\
0.80110 & 0.84453 & 0.88127 & 0.91217 & 0.93803 & 0.95958 \\
0.97747 & 0.99223 & 1.00000 & & &
\end{bmatrix},$$

$$\begin{bmatrix}
1.00000 & 2.69722 & 5.17136 & 8.89455 \\
14.24790 & 22.07157 & 33.61283 & 50.48096 \\
74.79993 & 109.40055 & 158.00460 & 225.44108 \\
317.89560 & 443.19350 & 611.11558 & 833.74392 \\
1125.83405 & 1505.20759 & 1993.15829 & 2614.86254 \\
3399.78420 & 4382.06240 & 5600.87014 & 7100.73115 \\
8931.78242 & 11149.97021 & 13817.16841 & 17001.20930 \\
20775.81856 & 23967.33875 & 25527.64563 & 25671.22552 \\
24609.29622 & 22640.51220 & 20147.13482 & 17477.63530 \\
14859.86462 & 12414.92533 & 10201.44191 & 8241.50255 \\
6534.43202 & 5066.17865 & 3815.60705 & 2758.60264 \\
1870.64631 & 1128.33931 & 510.47983 & 0.00000 \\
0.00000 & & &
\end{bmatrix}.$$

## Appendix C
## Runaway Latitude–Longitude Maps

Figure 5 shows the surface maps of temperature, precipitation, and evaporation at ISR = 1580 W m$^{-2}$, beyond the runaway installation threshold.

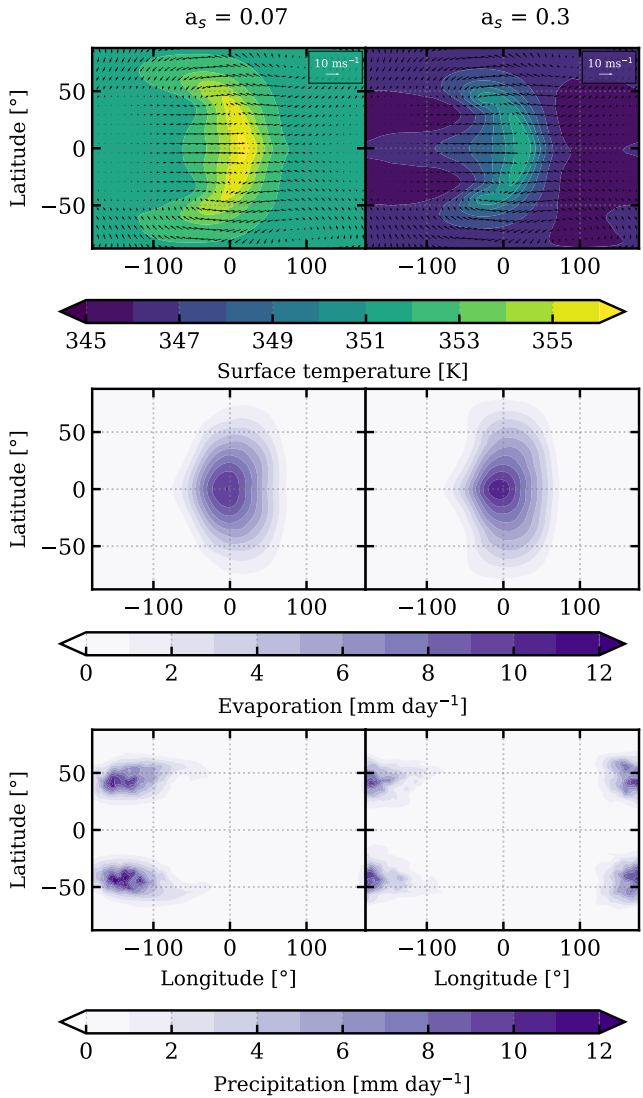

**Figure 5.** Latitude–longitude maps of surface temperature overlaid with the wind field at the bottom layer, and surface precipitation, for ISR = 1580 W m$^{-2}$, which brings the climate to the onset of a runaway greenhouse effect. The minimum surface temperature between the two surface albedo cases is 345 K or 72°C.

### ORCID iDs

Ryan Boukrouche ● https://orcid.org/0000-0002-5728-5129
Rodrigo Caballero ● https://orcid.org/0000-0002-5507-9209
Neil T. Lewis ● https://orcid.org/0000-0002-3724-5728